\documentstyle[12pt,epsf]{article}

\newcommand{\vm}[1]{\mbox{\bf#1}}
\newcommand{\vms}[1]{\mbox{\scriptsize{\bf#1}}}
\def\pj{\hspace{-.26cm}}
\def\fpj{\hspace{-.7cm}}
\def\thalf{{\textstyle{\frac{1}{2}}}}
\title{\begin{flushright}
{\normalsize NUC-MINN-98/3-T\\
July 1998 \\}
\end{flushright}
\vspace*{0.1in}
{\bf LOW-ENERGY THEOREMS FOR GLUODYNAMICS AT FINITE TEMPERATURE}}
\author{{\bf P. J. Ellis, J. I. Kapusta and H.-B. Tang}\\
{\it School of Physics and Astronomy}\\
{\it University of Minnesota}\\
{\it Minneapolis, MN 55455}}
\date{~}

\begin{document}

\maketitle

\begin{abstract}
We generalize the low-energy theorems of gluodynamics to finite temperature.
Examples of the theorems in the low and high temperature limits are given.
\end{abstract}

\newpage

The pure glue sector of quantum chromodynamics possesses scale and
conformal invariance at the classical level. Although this is destroyed by
quantum effects which introduce a scale, its residual effect does allow
low-energy theorems to be derived \cite{novikov81}.
These are important constraints on an effective theory.
In particular, they require a unique form for the effective Lagrangian
of a scalar glueball field \cite{migdal82}. The purpose of the present
note is to investigate how these relations are modified at finite
temperature, $T>0$.

In the imaginary time approach the partition function of gluodynamics is
\cite{kap}
\begin{equation}
Z=\int[d\bar{A}_a^\mu]\exp\left(-\frac{1}{4g_0^2}\int\limits_0^{1/T} d\tau
\int d^3x\:\bar{F}_a^{\mu\nu}\bar{F}^a_{\mu\nu}\right)\;.
\end{equation}
Here the gluon fields and field strength tensors have been
scaled by the bare coupling constant $g_0$: $\bar{A}_a^\mu=g_0{A}_a^\mu$
and $\bar{F}^a_{\mu\nu}=g_0{F}^a_{\mu\nu}$. For brevity we suppress the
gauge fixing and Faddeev-Popov ghost terms. Defining the grand potential
in the usual way, $\Omega=-T\ln Z$, one has
\begin{equation}
\frac{\partial}{\partial\left(-{8\pi^2}/{bg_0^2}\right)}\frac{\Omega}{V}
=-\frac{bg_0^2}{32\pi^2}\langle{F}_a^{\mu\nu}(0,\vm{0})
{F}^a_{\mu\nu}(0,\vm{0})\rangle\;,\label{eq:do}
\end{equation}
where $V$ is the volume of the system and $b=11N/3$ in pure
gluodynamics with the SU(N) gauge group.
The angle brackets signify a thermal average and we use the fact
that it is independant of position and imaginary time.

In making a dimensional analysis of the intensive quantity $\Omega/V$
we must acknowledge the existence of the dimensional parameter which
is generated by quantum effects.  At one-loop order it is
\begin{equation}
\Lambda=M_0\exp\left(-\frac{8\pi^2}{bg_0^2}\right)\;,
\label{eq:lam}
\end{equation}
where $M_0$ is the mass of the ultraviolet regulator.
At finite temperature we have an additional
dimensionful parameter, namely,
the temperature $T$. It follows that the grand potential per unit volume
must take the form
\begin{equation}
\frac{\Omega}{V}=\Lambda^4f(\Lambda/T)\;,\label{eq:oph}
\end{equation}
where $f$ is an unknown function.
It is sufficient to note that at zero temperature a form
$\propto\Lambda^4$ can be formally justified within a well-defined
regularization scheme \cite{novikov81}.
From (\ref{eq:oph}) one obtains
\begin{equation}
\frac{\partial}{\partial\left(-{8\pi^2}/{bg_0^2}\right)}\frac{\Omega}{V}
=\left(4-T\frac{\partial}{\partial T}\right)\frac{\Omega}{V}\;.
\label{eq:doph}
\end{equation}
Using Eqs. (\ref{eq:do}) and (\ref{eq:doph}) we have
\begin{equation}
\left(4-T\frac{\partial}{\partial T}\right)\frac{\Omega}{V}
=\frac{\beta_s(\alpha_s)}{4\alpha_s}\langle{F}_a^{\mu\nu}(0,\vm{0})
{F}^a_{\mu\nu}(0,\vm{0})\rangle=
\langle\theta_\mu^\mu(0,\vm{0})\rangle={\cal E} -3P\;.
\label{eq:basic}
\end{equation}
Here we have used standard thermodynamics to invoke the relationship
among the ensemble average of the trace of the energy-momentum
tensor density $\theta_\mu^\mu$, the energy density ${\cal E}$,
and the pressure $P$.  While we have only discussed
the first term of the Gell-Mann-Low function $\beta_s(\alpha_s)$,
there is no difficulty in including the
additional terms present in the complete $\beta_s$-function.

Differentiating Eq. (\ref{eq:do}) $n$ times and using (\ref{eq:doph})
we derive the low-energy relations
\begin{eqnarray}
&&\fpj (-1)^n \left(4-T\frac{\partial}{\partial T}\right)^{n+1}
\frac{\Omega}{V}=\left(T\frac{\partial}{\partial T}-4\right)^n
\langle\theta_\mu^\mu\rangle = \nonumber\\
&&\fpj \int d\tau_nd^3x_n\cdots\int d\tau_1d^3x_1
\left\langle\theta_\mu^\mu(\tau_n,\vm{x}_n)\cdots
\theta_\mu^\mu(\tau_1,\vm{x}_1)\theta_\mu^\mu(0,\vm{0})\right\rangle_c\;.
\label{eq:nglu}
\end{eqnarray}
Here the subscript $c$ indicates that only connected diagrams are to be
included. The limits of the imaginary time integration, 0 and $1/T$,
are suppressed here and henceforth.
A similar analysis for an operator ${\cal O}$ of dimension [mass]$^d$ gives
\begin{equation}
\left(T\frac{\partial}{\partial T} -d \right)^n
\langle{\cal O}\rangle=
\int d\tau_nd^3x_n\cdots\int d\tau_1d^3x_1
\left\langle\theta_\mu^\mu(\tau_n,\vm{x}_n)\cdots
\theta_\mu^\mu(\tau_1,\vm{x}_1){\cal O}(0,\vm{0})\right\rangle_c\;.
\label{eq:ngluop}
\end{equation}
As in \cite{novikov81} we assume that the operator here is constructed from
the gluon fields such that additional scales are
not introduced and that at zero temperature
$\langle{\cal O}\rangle$ vanishes in perturbation theory so that the
cut-off $M_0$ only appears in the combination (\ref{eq:lam}).

We can also generalize the finite momentum low-energy theorems of
Migdal and Shifman \cite{migdal82}. In imaginary time we define
\begin{eqnarray}
\Pi_2(\omega_m,\vm{p})&\pj=&\pj
\int d\tau_1d^3x_1 e^{-i(\omega_m\tau_1+\vms{p}\cdot\vms{x}_1)}
\left\langle{\cal O}(\tau_1,\vm{x}_1)
{\cal O}(0,\vm{0})\right\rangle_c\nonumber\\
\Pi_n(\omega_m,\vm{p})&\pj=&\pj\int d\tau_{n-1}d^3x_{n-1}\cdots
\int d\tau_1d^3x_1 e^{-i(\omega_m\tau_1+\vms{p}\cdot\vms{x}_1)}
\nonumber\\
&&\qquad\qquad\left\langle\theta_\mu^\mu(\tau_{n-1},\vm{x}_{n-1})\cdots
\theta_\mu^\mu(\tau_2,\vm{x}_2){\cal O}(\tau_1,\vm{x}_1)
{\cal O}(0,\vm{0})\right\rangle_c\;,
\end{eqnarray}
where $\omega_m$ denotes the Matsubara frequency and $n>2$.
Clearly
\begin{equation}
\frac{\partial^{n}\Pi_2}{\partial\left({8\pi^2}/{bg_0^2}
\right)^{n}}=\Pi_{2+n}\;.\label{eq:dnpi}
\end{equation}
In general $\Pi_2$ will depend on $\omega_m$ and $p=\sqrt{\vm{p}^2}$
as well as $\Lambda$ and $T$, so on dimensional grounds we must have
the functional form
\begin{equation}
\Pi_2=\Lambda^{2d-4}g(T/\Lambda,p/\Lambda,\omega_m/\Lambda)\;,
\end{equation}
where $g$ is an unknown function. Now the Matsubara frequency is a
discrete variable and so it is simplest to discuss the case $\omega_m=0$
(static correlation functions).  Then one deduces that
\begin{equation}
\left(p\frac{\partial}{\partial p}+T\frac{\partial}{\partial T}
+4-2d\right)^n\Pi_2=\Pi_{2+n}\;,\label{eq:nglupi}
\end{equation}
using (\ref{eq:dnpi}) and remembering to evaluate $\Pi_2$ and 
$\Pi_{2+n}$ at $\omega_m=0$.

It is known that pure SU(N) gauge theory has a second order phase
transition for N=2 and a first order transition for N=3 \cite{lattice}.
Therefore it is interesting to study these low-energy theorems at
low temperatures, where glueballs are the relevant degrees of
freedom, and at high temperatures, where the system is an interacting
plasma of quasi-free gluons.

At low temperatures  we wish to satisfy
Eq. (\ref{eq:nglu}) with a scalar glueball field $\phi$. We write the
effective Lagrangian
\begin{equation}
{\cal L}=\thalf\partial_\mu\phi\partial^\mu\phi-B_0
\,H\!\left(\frac{\phi}{\phi_0}\right)\;,
\end{equation}
where $B_0$ and $\phi_0$ are constants of mass dimension 4 and 1, 
respectively, and $H$ is an unknown function. The
trace of the energy-momentum tensor is
\begin{equation}
\theta_\mu^\mu=\left(4-\phi\frac{\partial}{\partial\phi}\right)B_0\,
H\!\left(\frac{\phi}{\phi_0}\right) =
D\,B_0\,H\left(\frac{\phi}{\phi_0}\right)\;,
\end{equation}
where $D = \phi_0\partial/\partial\phi_0 + 4B_0\partial/{\partial}B_0$,
and the partition function takes the standard form 
$Z=\int[d\phi]\exp\left(\int d\tau d^3x{\cal L}\right)$.
In order to obtain the structure of Eq. (\ref{eq:nglu}), namely
\begin{equation}
(-1)^nD^{n+1}\frac{\Omega}{V}=\int d\tau_nd^3x_n\cdots\int 
d\tau_1d^3x_1\left\langle\theta_\mu^\mu(\tau_n,\vm{x}_n)\cdots
\theta_\mu^\mu(\tau_1,\vm{x}_1)\theta_\mu^\mu(0,\vm{0})\right
\rangle_c\;,\label{eq:Dnpo}
\end{equation}
it is necessary that $D\theta_\mu^\mu=D^2B_0H(\phi/\phi_0)=0$. This
differential equation is easily solved, yielding $H(z)=z^4$ or
$z^4\ln z$. Thus one obtains the usual potential which breaks scale
invariance,
\begin{equation}
B_0\,H\!\left(\frac{\phi}{\phi_0}\right)=B_0\left(\frac{\phi}{\phi_0}\right)^4
\left[\ln\frac{\phi}{\phi_0}-\frac{1}{4}\right]\;,
\end{equation}
with $\phi_0$ denoting the $T=0$ vacuum field. On dimensional grounds
\begin{equation}
\frac{\Omega}{V}=B_0\,F\!\left(\phi_0/T, B_0^{\frac{1}{4}}/T\right)\;,
\end{equation}
where $F$ is an unknown function. Then it is straightforward to show that
\begin{equation}
D^{n+1}\frac{\Omega}{V}=
\left(4-T\frac{\partial}{\partial T}\right)^{n+1}\frac{\Omega}{V}\;,
\end{equation}
which, together with Eq. (\ref{eq:Dnpo}), indicates that 
Eq. (\ref{eq:nglu}) is satisfied at finite temperature with the
standard form of the glueball potential.

It is instructive to verify Eq. (\ref{eq:nglu}) (for $n=0$) with the
phenomenological glueball Lagrangian. Writing the field
$\phi=\phi_0+\delta\phi$, and expanding out the fluctuations $\delta\phi$,
the glueball mass is given by $m_G^2=4B_0/\phi_0^2$ and the thermal part of
the unperturbed grand potential is
\begin{equation}
\left(\frac{\Omega}{V}\right)_{T}=T\int\frac{d^3p}{(2\pi)^3}
\ln\left(1-e^{-\omega/T}\right)\;,
\label{eq:unp}
\end{equation}
and
\begin{equation}
\left(4-T\frac{\partial}{\partial T}\right)
\left(\frac{\Omega}{V}\right)_{T}=m_G^2{\cal D}(0,\vm{0})
\label{eq:dunp}\;,
\end{equation}
where $\omega^2=p^2+m_G^2$.  We have defined the propagator \cite{kap}
\begin{equation}
{\cal D}(\tau,\vm{x})= T\sum_m\int\frac{d^3p}{(2\pi)^3}
\frac{1}{\omega_m^2+p^2+m_G^2}e^{i(\omega_m\tau+\vms{p}\cdot\vms{x})}\;,
\end{equation}
so that
\begin{eqnarray}
&&\fpj {\cal D}(0,\vm{0})=\int\frac{d^3p}{(2\pi)^3}
\frac{1}{\omega(e^{\omega/T}-1)}\,,\nonumber\\
&&\fpj\int d\tau d^3x
{\cal D}(\tau,\vm{x})=m_G^{-2}\,,\nonumber\\
&&\fpj \int d\tau d^3x{\cal D}(\tau,\vm{x})^2=-\frac{\partial
{\cal D}(0,\vm{0})}{\partial m_G^2} \;.
\end{eqnarray}
Now $\theta_\mu^\mu=-B_0\left(1+{\delta\phi}/{\phi_0}\right)^4$,
and the diagrams that can contribute to
$\langle\theta_\mu^\mu(0,\vm{0})\rangle$ are shown in Fig. 1(a).
They give
\begin{equation}
\langle\theta_\mu^\mu(0,\vm{0})\rangle=\left({\textstyle\frac{5}{2}}
-{\textstyle\frac{3}{2}}\right)m_G^2{\cal D}(0,\vm{0})\;,
\end{equation}
in agreement with Eq. (\ref{eq:dunp}). We can also examine the contribution
of Fig. 1(b) to the thermal part of the grand potential:
\begin{equation}
\left(\frac{\Omega}{V}\right)_{T}=
-\frac{25m_G^2}{8\phi_0^2}{\cal D}(0,\vm{0})^2\;,
\end{equation}
and
\begin{equation}
\left(4-T\frac{\partial}{\partial T}\right)
\left(\frac{\Omega}{V}\right)_{T}=-\frac{25m_G^2}
{4\phi_0^2}{\cal D}(0,\vm{0})\left[2{\cal D}(0,\vm{0})
-T\frac{\partial {\cal D}(0,\vm{0})}{\partial T}\right]
\label{eq:dfir}\;.
\end{equation}
The corresponding diagrams for $\langle\theta_\mu^\mu\rangle$
are shown in Fig. 1(c). The first of these arises from a single
$\delta\phi$ and three three-point vertices
from ${\cal L}$. Its contribution is
\begin{eqnarray}
&&\fpj\frac{125m_G^8}{8\phi_0^2}{\cal D}(0,\vm{0})^2\int d\tau_1d^3x_1
d\tau_2d^3x_2d\tau_3d^3x_3\nonumber\\
&&\qquad\times {\cal D}(\tau_1,\vm{x}_1)
{\cal D}(\tau_1-\tau_2,\vm{x}_1-\vm{x}_2)
{\cal D}(\tau_1-\tau_3,\vm{x}_1-\vm{x}_3)
=\frac{125m_G^2}{8\phi_0^2}{\cal D}(0,\vm{0})^2\;.\label{eq:one}
\end{eqnarray}
Proceeding in similar fashion, the remaining contributions to
$\langle\theta_\mu^\mu\rangle$ are found to be
\begin{eqnarray}
&&\fpj\frac{m_G^2}{8\phi_0^2}\biggl[
-250{\cal D}(0,\vm{0})m_G^2\frac{\partial {\cal D}(0,\vm{0})}
{\partial m_G^2}-110{\cal D}(0,\vm{0})^2-75{\cal D}(0,\vm{0})^2
\nonumber\\
&&\hspace{4.5cm}+150{\cal D}(0,\vm{0})m_G^2\frac{\partial
{\cal D}(0,\vm{0})}
{\partial m_G^2}+60{\cal D}(0,\vm{0})^2\biggr]\;.\label{eq:two}
\end{eqnarray}
Using the relation
\begin{equation}
2m_G^2\frac{\partial {\cal D}(0,\vm{0})}{\partial m_G^2}=
2{\cal D}(0,\vm{0})-T\frac{\partial {\cal D}
(0,\vm{0})}{\partial T}\;,
\end{equation}
it is straightforward to verify that the sum of Eqs. (\ref{eq:one})
and (\ref{eq:two}) correctly gives Eq. (\ref{eq:dfir}).

Finally, consider very high temperatures where perturbation theory
is valid for most quantities of interest.  To two-loop order the
pressure is \cite{kap}
\begin{equation}
P=\frac{\pi^2}{45}(N^2-1)T^4 \left[1-\frac{5N}{4\pi}\alpha_s\right] \, .
\end{equation}\\
The strong coupling constant as a function of temperature is
\begin{equation}
\alpha_s(T) = \frac{2\pi}{b\ln(cT/\Lambda)}\;,
\end{equation}
where the numerical value of $c$ is irrelevant for our purposes.  
From this we compute
\begin{equation}
\langle \theta^{\mu}_{\mu}\rangle = {\cal E} - 3P =
T^5 \frac{d}{dT}\left(\frac{P}{T^4}\right) \;,
\end{equation}
and
\begin{equation}
\left(T\frac{d}{dT}-4\right)^n \langle \theta^{\mu}_{\mu}\rangle=
\frac{b}{72}\left(\frac{-b}{2\pi}\right)^n (n+1)! N(N^2-1)
\alpha_s^{n+2} T^4 \, .\label{eq:last}
\end{equation}
This gives us the value that should be obtained for the correlation 
functions on the right hand side of Eq. (7) in model calculations
or in lattice gauge theory.
Note that Eq. (\ref{eq:last}) is proportional to the
$(n+1)$'st power of the quantity $b$ which occurs in
the scale-breaking beta-function.
Without the quantum scale factor $\Lambda$ the
ensemble average of the trace of the energy-momentum tensor would
vanish.

In conclusion we have shown that at finite temperature the low-energy
theorems (\ref{eq:nglu}), (\ref{eq:ngluop}) and (\ref{eq:nglupi}) hold
provided that the operator $T\partial/\partial T$ is appropriately
included. The well-known zero-temperature glueball potential is unchanged
and can be used to study pure SU(N) gauge theory at low temperatures.
We also illustrated the theorems in the high temperature gluonic
plasma phase.  These theorems can be applied straightforwardly to
numerical simulations of lattice gauge theory.  Extensions to
QCD at finite baryon density are under investigation.

This work was supported in part by the US Department of Energy under grant
number DE-FG02-87ER40328.

\newpage

\begin{figure}
\centerline{\epsfxsize=14cm
\epsfbox{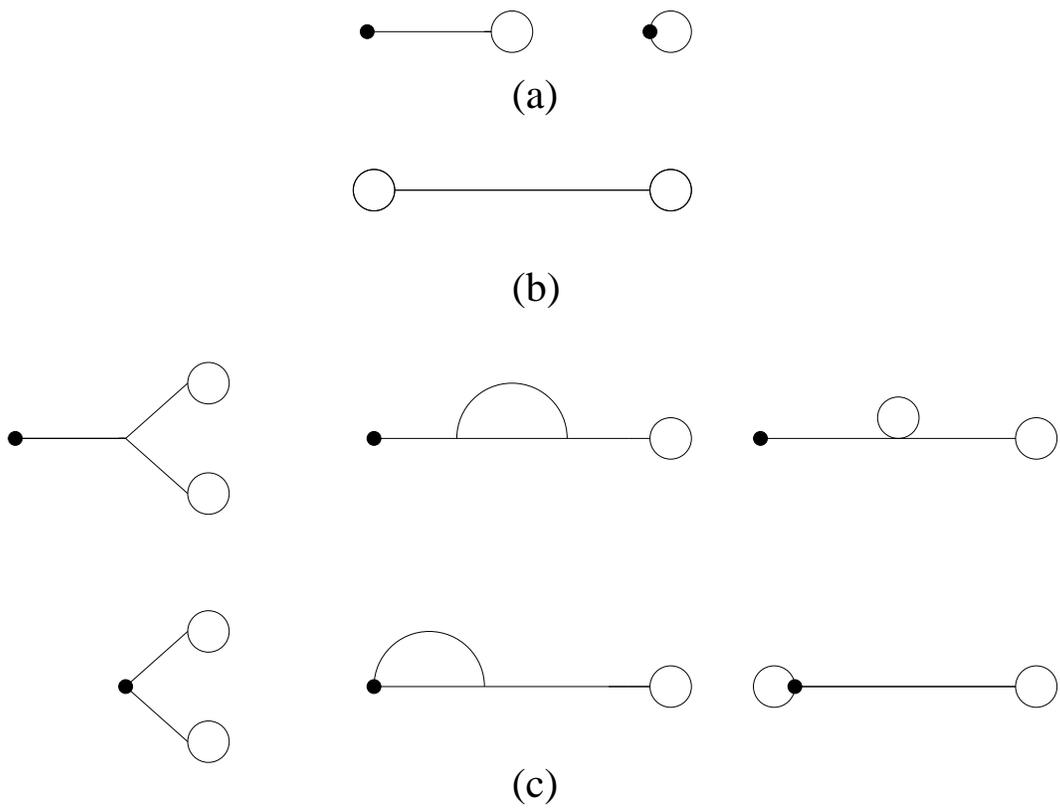}}
\caption{Diagrams for evaluating the identity Eq. (7), for $n=0$,
using an effective glueball Lagrangian at low temperature.
The solid dot represents the point $\tau=0$, ${\bf x}={\bf 0}$.}
\end{figure}

\begin{thebibliography}{99}
\bibliographystyle{unsrt}
\newcommand{\btem}{\bibitem}
\btem{novikov81} V. A. Novikov, M. A. Shifman, A. I. Vainshtein and
                 V. I. Zakharov, Nucl. Phys. {\bf B191} (1981) 301;
                 Sov. J. Part. Nucl. {\bf13} (1982) 224.
\btem{migdal82} A. A. Migdal and M. A. Shifman,
                Phys. Lett. {\bf B114} (1982) 445.
\btem{kap} J. I. Kapusta, {\it Finite Temperature Field Theory}
                (Cambridge University Press, 1985).
\btem{lattice} J. Fingberg, U. Heller and F. Karsch, Nucl. Phys.
               {\bf B392} (1993) 493.
\end{thebibliography}
\end{document}